# Ultrafast relaxation of hot phonons in Graphene-hBN Heterostructures

Dheeraj Golla, Alexandra Brasington, Brian J. LeRoy and Arvinder Sandhu*

*Department of Physics, University of Arizona, Tucson, United States*

ABSTRACT Fast carrier cooling is important for high power graphene based devices. Strongly Coupled Optical Phonons (SCOPs) play a major role in the relaxation of photoexcited carriers in graphene. Heterostructures of graphene and hexagonal boron nitride (hBN) have shown exceptional mobility and high saturation current, which makes them ideal for applications, but the effect of the hBN substrate on carrier cooling mechanisms is not understood. We track the cooling of hot photo-excited carriers in graphene-hBN heterostructures using ultrafast pump-probe spectroscopy. We find that the carriers cool down four times faster in the case of graphene on hBN than on a silicon oxide substrate thus overcoming the hot phonon (HP) bottleneck that plagues cooling in graphene devices.

Graphene heterostructures have garnered a lot of interest in the last decade[1]. Recently developed fabrication techniques have made it possible to engineer devices with better transport, optical and thermal properties[2,3]. Hexagonal boron nitride (hBN) is a layered material with a hexagonal lattice similar to graphene with a lattice constant that is about 1.8% larger[3]. It is an insulator with a wide band gap and high dielectric constant making it a good candidate as a substrate for graphene devices. Heterostructures of graphene and hBN show much higher mobility compared to those using $SiO_2$ as a substrate[2–4]. This improvement is a result of the hBN substrate being free of charged impurities and displacing the graphene away from the impurities in the $SiO_2$ substrate[4].

As electronic devices continue to scale down in size and push power capabilities, heat management has become a critical issue. Relaxation dynamics of photoexcited (PE) carriers has been studied extensively by many groups using variety of techniques such as photocurrent measurement, Raman time resolved Raman spectroscopy, transport measurements and ultrafast pump-probe spectroscopy[5–7] etc. Upon photoexcitation (with an ultrafast pulse for example), electrons and holes are excited, into a highly non-thermal system. This bath of carriers exchanges energy among themselves through coulombic interactions and thermalize into a hot (~1000's K) Fermi-Dirac population within tens of femtoseconds[8]. This hot thermal population cools further through the emission of optical phonons near the Γ point of the phonon dispersion. When the temperatures of the carriers and optical phonon bath equalize, this cooling channel slows down and this is termed as the Hot Phonon (HP) bottleneck[5,9–13]. Cooling through direct acoustic phonon emission is not viable because of a vanishingly small phase space for such a scattering process[14]. The hot optical phonons cool down through anharmonic decay to acoustic phonons which are subsequently absorbed into the substrate. Direct cooling of the charge carriers is also predicted to occur through coupling with the surface phonons of the underlying polar substrate[12,15–17]. Theoretical predictions and experiments place the hot optical phonon lifetime in graphene, graphite and CNTs in the 1-5ps range[5,9,18–21]. The buildup of optical phonons is detrimental to device performance and the HP bottleneck has been invoked to explain current saturation and negative differential conductance in graphene and CNTs[10,11,22]. The HP bottleneck also affects the photoresponse[23] of optoelectronic devices. It is important to explore cooling channels that can efficiently de-energize the optical phonons and remove the HP bottleneck. In that regard, graphene heterostructures incorporating an appropriate substrate, such as hBN, could offer additional mechanisms for accelerating the cooling process. It has been recently reported that the active cooling efficiency due to the Peltier effect in graphene-hBN devices is more than twice as much as the highest reported room temperature power factors[24]. A comparative study of relaxation dynamics for graphene on hBN and $SiO_2$ is missing from literature. In this letter, we study the relaxation of carriers in graphene-hBN heterostructure devices. Our findings indicate that the substrate interface plays a major role in the carrier cooling process and carriers in graphene devices fabricated on hBN substrates relax significantly faster than those on $SiO_2$ substrates thus providing relief of the HP bottleneck and enabling better device performance.

Hexagonal boron nitride flakes were exfoliated and deposited on silicon chips that have a 285 nm thermally grown oxide. Pristine graphene was grown on copper foil using a low pressure chemical vapor deposition (CVD) method as described in the work by Xuesong Li et al[25]. PMMA was spin coated onto the copper foil before floating it on a mixture of hydrogen peroxide, hydrochloric acid and de-ionized water to etch away the copper. The remaining graphene/PMMA film was transferred to clean de-ionized water. The $Si/SiO_2$ chip with exfoliated hBN was used to gently pick up the floating graphene/PMMA film and then dried. The chip was then placed in acetone to dissolve the PMMA layer. The samples were then subsequently annealed in an atmosphere of argon and hydrogen at 350˚C for 3 hours to get rid of residues, impurities and ensure better adhesion to the substrate. An optical image of one of the samples is shown in Figure 1. The spot marked 1 has graphene on hBN (g-hBN) whereas spot 2 has graphene on $SiO_2$ (g-SiO$_2$). The Raman spectra of g-SiO$_2$ and g-hBN are shown in Figure 1(b). The absence of a D peak means that both the g-SiO$_2$ and g-hBN are defect free. We infer from the G and 2D peak positions that g-SiO$_2$ and g-hBN are p-doped by about $3.5 \times 10^{12}$ $cm^{-2}$ and $1 \times 10^{12}$ $cm^{-2}$ respectively[26,27], which is well below the 1.58 eV probing photon energy.

For the pump-probe study we used amplified 780 nm pulses from a Ti-sapphire laser amplifier for both pump and probe. The spot sizes (FWHM) of the pump and probe beams were measured using the knife edge technique to be 154 μm and 23 μm respectively. The FWHM as measured using the FROG technique was 45 fs. The experiment was conducted with a range of pump pulse energies, all of which were below the damage threshold of graphene under irradiation with ultrafast pulses[28]. The pump was chopped using an optical chopper and the probe reflectivity of the sample was measured using lock-in detection. The polarizations of the pump and probe were crossed for better rejection of the pump scatter.

The relaxation of PE carriers is captured by the differential reflectivity, $\frac{\Delta R(t)}{R_0}$, of the sample. The differential reflectivity shows the opposite trend as the electronic temperature in graphene, which means that a decrease in $\frac{\Delta R(t)}{R_0}$ corresponds to an increase in the electronic temperature. Figure *2* shows $\frac{\Delta R(t)}{R_0}$ as a function of the pump-probe time delay for three different g-hBN structures along with curves for g-SiO$_2$ for comparison. The pump pulse energy is 60 $\frac{\mu J}{cm^2}$. The baseline at large positive pump probe delay is non-zero because of the underlying contribution from silicon base of our samples. This baseline is constant over 100 picoseconds which is far greater timescale than those discussed in this letter. We have independently verified that this baseline does not contribute to the lifetimes extracted from our experiment (section S2 in the supplementary information). It is immediately evident from Figure *2* that the relaxation dynamics of g-hBN are faster than that of g-SiO$_2$.

In order to quantify the timescales observed in the experiment we modeled the temperature dynamics of the heterostructure using a two-temperature model[29,30]. The lateral transport of heat is negligible because the diffusion timescale is of the order of 90 microseconds (section S1 in the supplementary information).

$$\frac{dT_{el}}{dt} = \frac{I(t) - \Gamma(T_{el}, T_{op})}{c_{el}(T)} \qquad (1)$$

$$\frac{dT_{op}}{dt} = \frac{\Gamma(T_{el}, T_{op})}{c_{op}(T)} - \frac{T_{op} - T_0}{\tau_{op}} \qquad (2)$$

where $T_{el}$ and $T_{op}$ denote the electronic and optical phonon temperatures of graphene respectively. The coupling between $T_{el}$ and $T_{op}$ is given by the function $\Gamma(T_{el}, T_{op})$. $T_0$ is the ambient room temperature; $c_{op}$ and $c_{el}$ denote the phononic and electronic heat capacity of graphene; $I(t)$ is the time profile of the pump pulse which is assumed to be a Gaussian with a FWHM of 45 fs. The thermal relaxation timescale is $\tau_{op}$ which denotes the optical phonon lifetime in graphene. We numerically solve the system of differential equations given above for the electronic temperature, which determines the optical conductivity (σ) of graphene as a function of time. We use the optical conductivity calculated in the previous step to determine the total reflectance of the heterostructure using the transfer matrix method. We fit the experimental transient reflectivity curves using the two temperature model to estimate the optical phonon relaxation lifetime $\tau_{op}$. Refer to the supplementary information (section S3) for a detailed explanation of the model. The results of the fitting process are shown in figure 3. Figure 3(a) shows the relaxation for g-hBN pumped with pulse fluences of 80 (orange), 60 (green) and 50 (blue) $\frac{\mu J}{cm^2}$ for which we extract optical phonon lifetimes of 375 fs, 250 fs and 200 fs ($\pm$ 25fs). Figure 3(b) shows the corresponding fits for g-SiO$_2$ and the corresponding lifetimes are 1500 fs, 1200 fs and 800 fs ($\pm$ 50fs). The inset shows the temperature profiles for the case with the fastest decay. We assume that the only cooling mechanism available to the electrons is through losing heat to the optical phonon bath. Other possible modes of electronic cooling, such as directly coupling to the substrate or coupling to the acoustic phonons, are neglected. The interplay between the electron-phonon coupling strength ($\beta$) and optical phonon lifetime ($\tau_{op}$) determines whether the electrons cool down substantially before equilibrating with the phonons, i.e. before the bottleneck sets in. The fast decay of optical phonons in the case of g-hBN has an immediate cooling effect on the electronic temperature. As seen in the temperature profile for g-SiO$_2$, T$_{el}$ and T$_{op}$ equalize around 1200K whereas for g-hBN the temperatures have cooled down to 600K before equalizing. The phonon relaxation lifetimes in g-hBN are lower than those measured for g-SiO$_2$ by about a factor of four for all fluences indicating that additional cooling channels for graphene's optical phonons are available when hBN is used as the substrate. The range of values of $\tau_{op}$ for g-SiO$_2$ (0.8-1.5ps) agrees well with previous measurements of the phonon lifetime in SiO$_2$ supported graphene[13,21]. At higher fluences (i.e. high initial carrier densities), multibody effects like carrier screening, plasmonic modes and plasmon-phonon interactions might come into play[31]. While these multibody effects are not captured by the simple two temperature model, it still allows us to phenomenologically deduce the optical phonon relaxation lifetime. At low fluence where these multibody effects are relatively small and the temperature differential is smallest, we can estimate the interfacial thermal conductance of the graphene – hBN interface from the optical phonon lifetimes using the lumped heat capacity model. The phonon lifetime is related to the interfacial conductance by the equation

$$G_k = \frac{c_{eff}}{\tau_{op}} \quad (3)$$

where, $\frac{1}{c_{eff}} = \frac{1}{c_{hBN}} + \frac{1}{c_{op}+c_{el}}$ (4)

The heat capacity $c_{eff}$ is the effective heat capacity per unit area of the composite graphene-hBN system[32]. The heat capacity and conductance of hBN can be ignored because the limiting term in the vertical heat transport dynamics of the heterostructure is the interfacial thermal conductance between graphene and the substrate. Since the measured phonon lifetime shows a decreasing trend with the temperature differential, we estimate the lower limit of the room temperature interfacial conductance of the graphene-hBN interface from our measured relaxation lifetime for the lowest fluence as 16.25 $\frac{MW}{m^2.K}$ at room temperature. The corresponding value for the graphene-SiO$_2$ interface is 3.75 $\frac{MW}{m^2.K}$. The value of $G_k$ for g-hBN measured here is higher than that reported in the work by Chen et al.[33] by more than a factor of two. The sample used in their experiment underwent electron beam lithography and oxygen plasma etching which might have possibly affected the interface quality and suppressed the interfacial conductance. As far as we know there are no previous measurements of the relaxation of carriers in graphene on hBN. For the purpose of comparison, we can calculate the equivalent relaxation times from the alternate methods for measurements and predictions of the interfacial thermal conductivity of the graphene – hBN interface. The results are listed in table 1.

The interaction between the carriers in graphene and the surface plasmon polaritons (SPP) of the polar substrate has been proposed as a possible cooling mechanism for overcoming the HP bottleneck in graphene[15–17,34]. It has been established that graphene on hBN substrates has lower charge doping level than graphene on SiO$_2$[3] which is also the case in our samples as evidenced by the slightly upshifted (~10 cm$^{-1}$) and narrower G peak[27] for g-SiO$_2$. If SPP interactions were the dominant cooling mechanism, the doping of graphene due to SiO$_2$ will shield this interaction and reduce the efficacy of this channel consequently increasing the relaxation time for phonons in g-SiO$_2$. The interaction between graphene and the substrate also depends on many factors like topographic conformity, coulombic interactions and adhesion energy. The g-hBN interface can be more transparent to heat carrying phonons because of the similar masses of carbon, boron and nitrogen[35]. The curvature of the graphene sheet is an additional contributor to the interface thermal resistance in g-SiO$_2$[36]. Annealing contributes to the graphene sheet conforming to the substrate and hBN being atomically flat means the graphene sheet in g-hBN has lower cumulative curvature than the graphene sheet in g-SiO$_2$ effectively decreasing interfacial resistance in g-hBN.

In conclusion, we have used differential reflectance spectroscopy to study the carrier dynamics of graphene-hBN heterostructures. We extract the optical phonon lifetime and interface thermal conductance using a two-temperature model. The thermal relaxation rates of graphene-hBN are significantly faster than those of graphene-SiO$_2$ thus mitigating the hot phonon bottleneck. We conclude that hBN substrates will enhance the thermal performance of high power graphene devices[39].

FIGURES

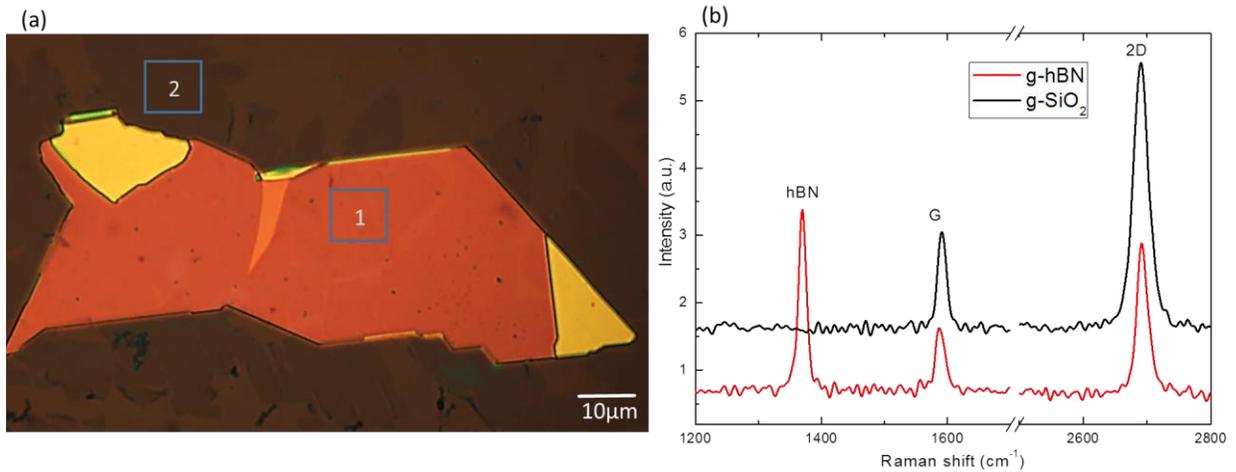

Figure 1. (a) Optical image of sample 1. The mean thickness of the hBN flake shown here is 118 nm as measured using AFM. Spot marked 1 is g-hBN and spot marked 2 is g-SiO$_2$ (b) Raman spectra of the graphene on hBN (red) and on SiO$_2$ (black). The curves are vertically offset for clarity.

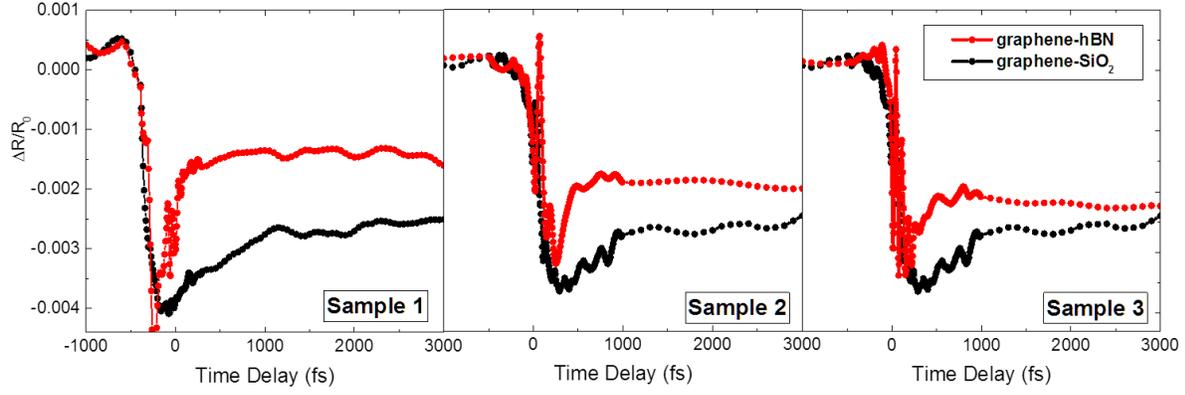

Figure 2. Experimental differential reflectivity curves for different samples of g-hBN pumped using $60\frac{\mu J}{cm^2}$ per pulse. The reflectivity of g-SiO$_2$ is shown for comparison (red: g-hBN and black: g-SiO$_2$)

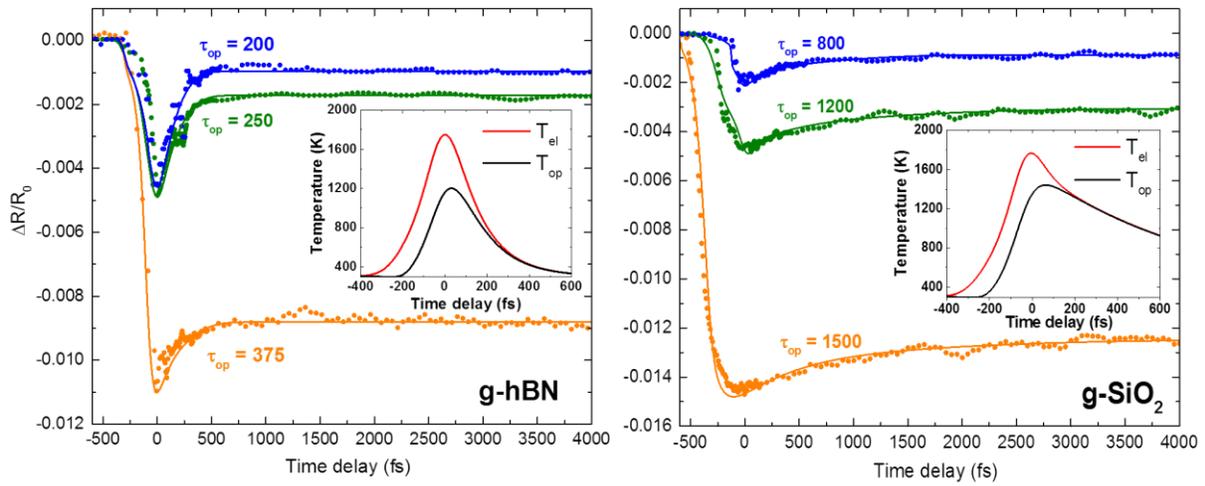

Figure 3. (a) Differential reflectivity curves for graphene-hBN with different pump fluences: $80\ \frac{\mu J}{cm^2}$ (orange), $60\ \frac{\mu J}{cm^2}$ (green) and 50 (blue) $\frac{\mu J}{cm^2}$. (b) Differential reflectivity curves for graphene-SiO$_2$ for the same fluences. Inset: Electronic (red) and phonon (black) temperature profiles for the case with the lowest fluence.

TABLES

Table 1. Interfacial thermal conductance and thermal time constants for graphene on hBN

| Study | $G_k \left(\frac{MW}{m^2 \cdot K}\right)$ | Thermal time constant (fs) | Notes |
|---|---|---|---|
| Mao, R. et al[35] | 187 | 17 | Room temperature, theoretical |
| Pak, A. J. et al[36] | 4 | 800 | Room temperature, theoretical |
| Chen, C. C. et al[33] | 7.41 | 435 | Room temperature, experimental |
| Zhang, J. et al[37] | 3 | 1076 | 200-700K, theoretical |
| Ting Li et al[38] | 1-10 | 300-3000 | 200-600, theoretical |

*Corresponding Author

asandhu@email.arizona.edu

SUPPLEMENTARY MATERIAL

S1. Lateral heat transport

S2. hBN thickness dependence and baseline of $\frac{\Delta R(t)}{R_0}$ curves

S3. Details of the two temperature model and reflectivity calculation

S4. Evolution of the electronic and phonon temperature

ACKNOWLEDGMENT

A. B. and B.J.L. were supported by the U.S. Army Research Laboratory and the U.S. Army Research Office under Contract/Grant No. W911NF-14-1-0653. This work was supported by the National Science Foundation (NSF) under contract PHY 1505556.

# Supplementary Information for "Ultrafast relaxation of hot phonons in Graphene-hBN Heterostructures"

**S1. Lateral Heat transport**

The lateral heat transport can be safely ignored because the spot size of the laser is large. The diffusion time for the heat to spread laterally in the region being pumped by the laser spot is

$$t = \frac{A}{\alpha} = \frac{2.25 \times 10^{-4} \, cm^2}{2.5 \, \frac{cm^2}{s}} = 90 \mu s$$

A is the area of the pump spot and α is the diffusivity of graphite[1]. Thus the lateral diffusion time is much too large compared to the timescale of vertical heat transport (~ 5ps) that is probed in the experiment.

**S2. Dependence on hBN thickness and non-zero baseline of $\frac{\Delta R(t)}{R_0}$ curves**

The addition of hBN flake does not change the response of silicon in any way since hBN is transparent to the pump light (780nm). The thickness of the hBN doesn't affect the relaxation timescale as seen from figure S2(a). The $\frac{\Delta R(t)}{R_0}$ versus time delay measurements shown in the manuscript relax to a non-zero value. This is the residual signal shown by the silicon chip on which the graphene and hBN flakes are deposited. The reflectivity curves for the plain Si/SiO$_2$ show that we can approximate the silicon response with a sigmoidal function. This response has no decay as we have verified over 70ps (figure S2(b)).

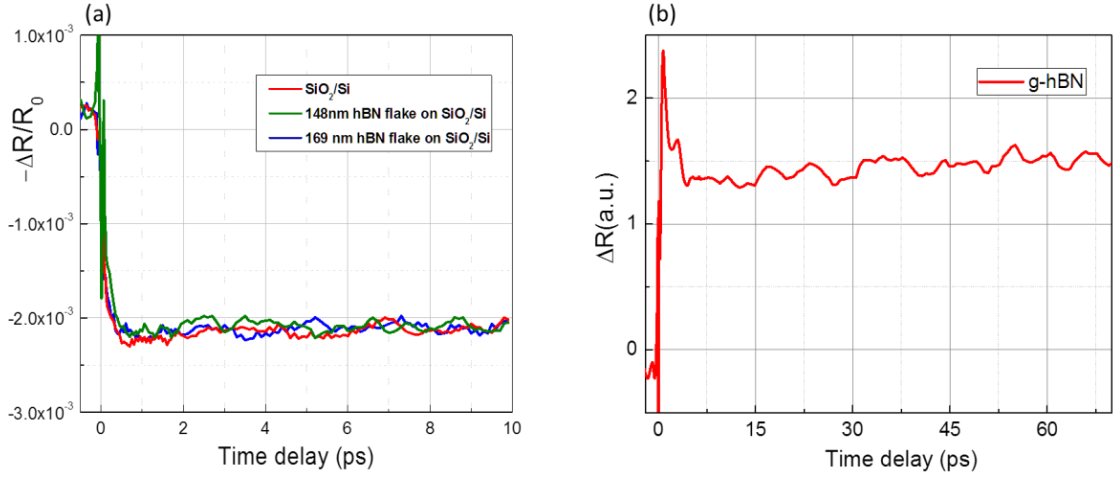

Figure S2. (a) Reflectivity curves for plain SiO$_2$/Si chip and hBN flakes over various thicknesses. (b) Long scan of time delay of g-hBN showing a flat response after the initial cooling.

**S3. Details of the two temperature model and reflectivity calculation**

The two temperature model makes the lumped heat capacity approximation wherein the limiting term in the heat dynamics is the interfacial thermal resistance. This is a reasonable approximation because the atomic layer thickness of graphene ensures a uniform temperature across its depth. Another factor contributing to the validation of the lumped element model is that the heat capacity per unit area of hBN is much higher than graphene.

$$\frac{dT_{el}}{dt} = \frac{I(t) - \Gamma(T_{el}, T_{op})}{c_{el}(T)}$$

$$\frac{dT_{op}}{dt} = \frac{\Gamma(T_{el}, T_{op})}{c_{op}(T)} - \frac{T_{op} - T_0}{\tau_{op}}$$

$$\Gamma(T_{el}, T_{op}) = \beta \left\{ \left(1 + n(\hbar\Omega, T_{op})\right) \int D(E) D(E - \hbar\Omega) f(E, T_{el})(1 - f(E - \hbar\Omega)) \, dE - n(\hbar\Omega, T_{op}) \int D(E) D(E + \hbar\Omega) f(E, T_{el})(1 - f(E + \hbar\Omega)) \, dE \right\}$$

where $T_{el}$ and $T_{op}$ denote the electronic and optical phonon temperatures of graphene respectively. $\Gamma(T_{el}, T_{op})$ describes the coupling between $T_{op}$ and $T_{el}$. The coupling constant (β) is taken as $8\frac{eV^2}{cm^2.s}$ which is the best fit for the experimental data. $D(E)$ is the density of electronic states of graphene is given by $D(E) = \frac{2E}{\pi}(\hbar v_f)^{-2}$. The Fermi-Dirac formula for the distribution of carriers is $f(E, T_{el}) = \frac{1}{\exp\left(\frac{E}{kT_{el}}\right) - 1}$. The Bose-Einstein occupation of optical phonons at energy $\hbar\Omega$ is $n(\hbar\Omega, T_{op}) = \frac{1}{\exp\left(\frac{\hbar\Omega}{kT_{op}}\right) + 1}$. The two optical phonons that contribute are the E$_{2g}$-mode phonons near the Γ-point and A$_1$-mode phonons near the K-point of graphene. We neglect the dispersion and assume that $\hbar\Omega \approx 200 meV$. The expressions for the $c_{el}$ and $c_{op}$ are derived from theory and experiment respectively and the values for which are taken from the work by Chun Hung Lui et al[2]. The graphene was assumed to be intrinsic after verifying that the doping only affects the maximum T$_{el}$ reached by 2% and doesn't change the time dynamics. We keep the electron-phonon interaction strength (β) fixed for all cases and vary the absorbed fluence to match the peak electronic temperature with the experimental value. The model can easily be extended to include the transient heating of the hBN (and for the SiO$_2$ underneath) due to the inflow of heat from the graphene layer but the rise in temperature of the substrate is negligible (~1K) because of the large specific heat capacity of hBN. Once the electronic temperature is obtained from the above equations, we can extract the optical conductivity (σ) and refractive index (η) of graphene from the equations:

$$\sigma_{inter}(T) = \frac{\pi e^2}{2h} \tanh\frac{E_{probe}}{4kT}$$

$$\sigma_{intra}(T) = \frac{\pi e^2}{2h}\left(\frac{8\ln(2)}{\pi}\right)\frac{\gamma kT}{(E_{probe}^2 + \gamma^2)}$$

$$\eta_{graphene} = \sqrt{\varepsilon_{core} + \frac{i(\sigma_{inter} + \sigma_{intra})}{\varepsilon_0 \omega}}$$

$\sigma_{inter}$ and $\sigma_{intra}$ denote the interband and intraband contributions to the optical conductivity respectively. The contribution of the core electrons (that don't participate in electronic transitions) to the relative permittivity of graphene is denoted as $\varepsilon_{core}$. We calculate the reflectivity of the graphene\hBN\SiO$_2$\Silicon stack using the transfer matrix method of thin film interference[3]. Thus, knowing the electronic temperature $T_{el}(t)$ as a function of time delay allows us to calculate reflectivity $R(t)$ as a function of time delay.

## S4. Evolution of the electronic and phonon temperature

The following graphs show the evolution of the electron and phonon temperatures extracted from the simulations with incident fluence 60 $\frac{\mu J}{cm^2}$ (Fig S4) and 80 $\frac{\mu J}{cm^2}$ (Fig S5) respectively. We note that the maximum electron temperature reached can vary non-linearly with the incident pump fluence due to state filling, localized doping or screening due to the substrate.

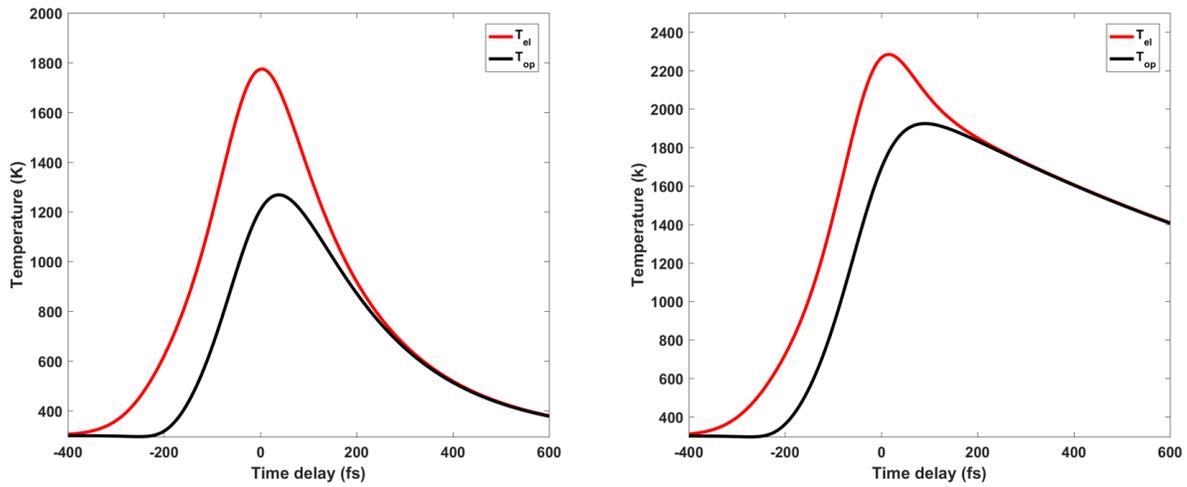

Figure S4. Temperature profile for pump fluence of $60 \frac{\mu J}{cm^2}$. Left: Temperature profiles for g-hBN with $\tau_{op} = 250 fs$. Right: Temperature profiles for g-SiO$_2$ with $\tau_{op} = 1200 fs$

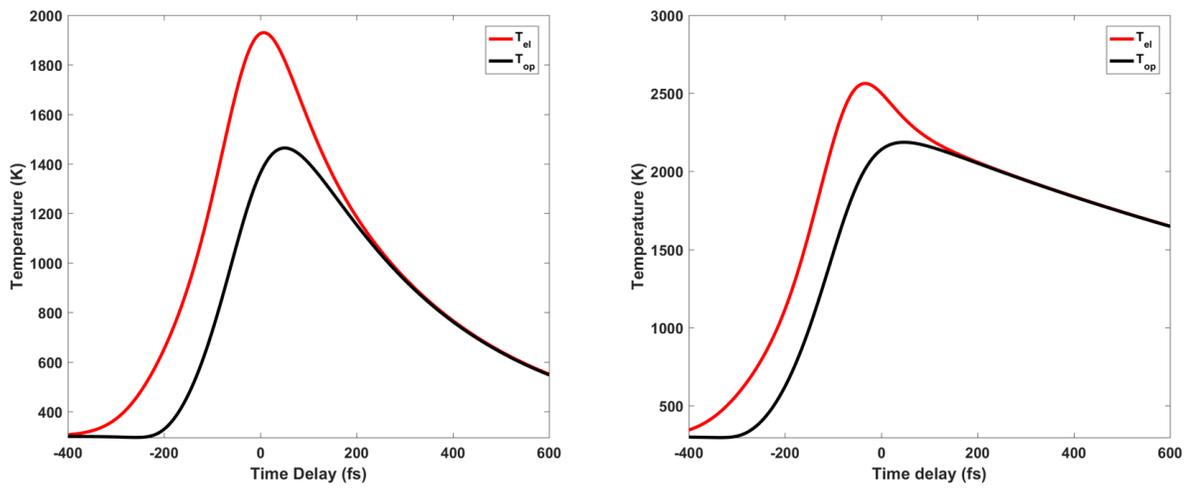

Figure S5. Temperature profile for pump fluence of $80 \frac{\mu J}{cm^2}$. Left: Temperature profiles for g-hBN with $\tau_{op} = 375 fs$. Right: Temperature profiles for g-SiO$_2$ with $\tau_{op} = 1500 fs$